# Assessing the Affordability of Nutrient-Adequate Diets

Kate R. Schneider[1], Luc Christiaensen[2], Patrick Webb[3], William A. Masters[3]

*Last revised: 14 July 2022*

## Abstract

The cost and affordability of least-cost healthy diets by time and place are increasingly used as a proxy for access to nutrient-adequate diets. Recent work has focused on the nutrient requirements of individuals, although most food and anti-poverty programs target whole households. This raises the question of how the cost of a nutrient-adequate diet can be measured for an entire household. This study identifies upper and lower bounds on the feasibility, cost, and affordability of meeting all household members' nutrient requirements using 2013-2017 survey data from Malawi. Findings show only a minority of households can afford the nutrient-adequate diet at either bound, with 20% of households able to afford the (upper bound) shared diets and 38% the individualized (lower bound) diets. Individualized diets are more frequently feasible with locally available foods (90% vs. 60% of the time) and exhibit more moderate seasonal fluctuation. To meet all members' needs, a shared diet requires a more nutrient-dense combination of foods that is more costly and exhibits more seasonality in diet cost than any one food group or the individualized diets. The findings further help adjudicate the extent to which nutritional behavioral change programs versus broader agricultural and food policies can be relied upon to improve individual access to healthy diets.

[1] School of Advanced International Studies, Johns Hopkins University; [2] World Bank; [3] Friedman School of Nutrition Science & Policy, Tufts University

**Introduction**

Whether markets can provide access to affordable, nutritious diets year-round is an important criterion for assessing national and global food system performance (FAO et al. 2020). Least-cost diets for individuals, based on recommended nutrient intakes and obtained through linear programming, are increasingly used as an operational metric of access to foods with adequate nutrients for an active and healthy life.[1] This study extends that work to whole households, because agricultural policies or social safety nets typically target the entire household, and because foods are often procured and prepared for the household as a whole and then shared among household members to varying degrees (Berti 2012). Evaluating whether households have access to nutrient-adequate diets requires considering the nutrient needs of each individual household member, and then addressing whether they consume individually tailored or shared meals (Schneider et al. 2021).

To measure each household's access to nutrient-adequate diets, one option would be to solve for every individual member's least-cost combination of foods and add them up, while the opposite benchmark is to solve for a single shared diet that meets all members' needs. Actual households might not be able to prepare and serve a different diet to each person, but the thought experiment reveals the lower bound on food costs to meet all members' nutrient needs. Similarly, the corresponding upper bound would be a shared diet that uses one (smaller) set of foods to meet all members' needs, as recently developed in the nutrition literature (Schneider et al. 2021). By construction, the nutrient-adequate shared diet must have the same or greater nutrient density than the sum of the individualized diets, because the shared diet's level of each vitamin and mineral is that of the household's neediest member (Schneider et al. 2021). For example, a household with one man and one woman would need more iron per calorie in its shared diet than with two individualized diets, because women need a higher level of iron per calorie (Bai, Herforth and Masters 2022). The present study measures how these demographic differences in nutrient requirements among surveyed households affect the feasibility, cost, and affordability of meeting nutrient needs, given the nutrient composition and prices of foods available at local markets in Malawi from 2013 to 2017.

This paper is part of a trio, preceded and followed by closely related studies, and situated within a larger body of work on the affordability of healthy diets globally (Bai et al. 2022; Masters et al. 2018; Herforth et al. 2020). It builds on Schneider et al. (2021) which developed the shared household nutrient requirements, compared shared to individual nutrient requirements, and assessed the adequacy of observed diets relative to the shared requirements. Findings show diets in Malawi, on average, are lacking in most nutrients and unbalanced in macronutrients. Meeting the needs of members with a shared household diet would require much more iron, zinc, phosphorus, and vitamin C than required by individuals alone. The present study is followed by Schneider (2022), which uses a combination of shadow price and regression analyses with policy simulations to deduce the drivers of an infeasible least-cost shared diet. Drivers of infeasibility are found to be household composition and incompatible presence of selenium and copper in foods, preventing a diet that satisfies the former within the limits of the latter.

---

[1] Different nutrition criteria have been applied, including energy adequacy (Herforth et al. 2020; FAO et al. 2020); minimum scientific nutrient requirements for optimal growth and long-term health (Bai et al. 2021; Bai et al. 2022; Masters et al. 2018; Herforth et al. 2020); food-based dietary guidelines (Raghunathan, Headey and Herforth 2021; Mahrt et al. 2019; Dizon, Herforth and Wang 2019; Herforth et al. 2020); and the EAT-Lancet sustainable diet recommendations (Hirvonen et al. 2019; Willett et al. 2019).

The cost and affordability of nutrient-adequate diets depends not only on nutrient requirements, but also on which foods are available at what prices in each location and time. In many rural settings across the world, and especially in Africa, the availability and cost of individual foods changes substantially across seasons (Bai, Naumova and Masters 2020; Gilbert et al. 2017; Kaminski, Christiaensen and Gilbert 2016). As shown by Bai et al. (2020), substitution among foods to meet individual nutrient requirements can moderate seasonal fluctuation in diet costs only to a limited degree, because of synchronized fluctuations in availability and price of items in the same food group that have similar nutrient composition (Arimond et al. 2010; Fiedler and Lividini 2017).

This paper quantifies the implications of demographic heterogeneity for households' monthly access to nutrient-adequate foods in rural Malawi, as an example for similar analyses that might be conducted elsewhere. We make two main contributions to the literature on access to nutritious diets. The first contribution is methodological; we extend the least-cost diets framework from individuals to households building on recently developed shared nutrient requirements (Schneider et al. 2021). Our second contribution is empirical, showing the nature and magnitude of variation in diet costs based on nationally representative household survey data, linking information about household composition with sub-national monthly food prices and country-specific food composition data. Both contributions rely on using both individualized and shared diets to provide lower and upper bounds on the cost of nutrient-adequate foods that meet the needs of each surveyed household member, using the food items and prices reported for the district market in each month.

Providing a range of diet costs between the lower and upper bounds offers a more useful measure of food access than just one point estimate, as policies and programs could aim to help households with varying degrees of meal sharing. The feasibility, cost, and affordability of nutrient-adequate diets identified in this study measures access to the minimal set of foods needed to meet nutrient requirements, as a necessary but not sufficient condition for food security and adequate nutrition. Importantly, the cost of nutrient adequacy per day is lower than the cost to meet national dietary guidelines, because those require larger quantities of relatively expensive food groups (Herforth et al. 2020).[2] Capturing additional dimensions of food system performance would require an even greater range of data and computational complexity and could be addressed in future studies of individual and household consumption.

Access to adequate nutrients in rural Malawi is of particular interest because the country has made substantial progress towards improved nutrition, but continues to have among the world's highest levels of certain micronutrient deficiencies (National Statistical Office (NSO) Community Health Sciences Unit (CHSU) [Malawi], Centers for Disease Control and Prevention (CDC) and Emory University 2017). Using new methods that account for sharing within households is relevant everywhere that meals are eaten together "family style" (Hjertholm et al. 2019; Gelli et al. 2020; Hjertholm et al. 2018), but is specifically important in Malawi where households are relatively large with a wide range of ages eating together, often from a single plate. Finally, Malawi provides a useful case study because we can match two rounds of household surveys with monthly observations of market prices for many years, and thereby

---

[2] Furthermore, these diet costs refer only to the acquisition of food items without accounting for the time and materials needed to cook and prepare meals and refer only to the nutritional requirements for physical health without accounting for culinary practices, individual preferences, or cultural appropriateness. Nutrition security beyond food security further requires adequate care practices, sufficient sanitation, and good hygiene (United Nations Children's Fund (UNICEF) 1990).

address temporal variation and seasonality in diet costs.

With maize playing an outsized role in Malawi's food policy, as well as in consumers' diets, and maize prices in Malawi displaying particularly high seasonality (Sibande, Bailey and Davidova 2017; Gilbert et al. 2017; Schneider et al. 2021; Pauw et al. 2018), the empirical findings from Malawi regarding seasonality in the cost of adequate diets are also of particular interest. As indicated above, it is not clear *a priori* whether the cost of whole diets will follow similar seasonal trends identified in studies of single food items or food groups. First, nutritionally adequate whole diets require a combination of foods whose seasonality patterns may differ in periodicity, for example, several nutrient-dense foods are harvested during the rainy (lean) season (Chikhungu and Madise 2014; Gelli et al. 2020; Gilbert et al. 2017; Bai et al. 2020). Second, maize will play a smaller role in the adequate diet than it does in current (largely inadequate) diets, so maize prices may not drive the cost pattern.

The paper proceeds as follows. Section 2 elaborates the conceptual framework motivating and describing the two approaches to estimating household diet costs. Section 3 presents the data sources as well as key background features of Malawi's food system. Section 4 explains the methods used to define the nutrient requirements and calculate diet costs; estimate the extent of seasonality in diet costs; and assess affordability and overall access. Section 5 discusses the findings, followed by section 6 which concludes.

**Conceptual Framework**

Estimating a least-cost diet requires a set of foods, their prices, and establishing the constraints that a combination of foods must meet at the lowest total cost. For a nutrient-adequate diet, those constraints are the nutrient requirements, defined by lower bounds for almost all nutrients (excepting retinol and sodium) as well as upper bounds for some. These requirements differ by age, sex, maternity, and activity level. The lower bound on the household diet cost ("individualized diets") uses individual nutrient requirements and estimates separate least-cost diets for every person in the household. The upper bound ("household sharing") uses shared nutrient requirements, that are defined per household based on its composition. Shared requirements ensure that the resulting diet can meet total energy needs for the whole family and will have the nutrient density (nutrient quantity per calorie) that meets the minimum needs of the member with the greatest need (and not exceed the upper limit of the most sensitive member) when every member eats (just) enough of the shared diet to meet their energy needs.[3] Comparing current diets to this shared benchmark, Schneider et al. (2021) found nearly all households consuming diets lacking in riboflavin, selenium, lipids, or vitamin B-12 and 80% consuming more calories from carbohydrates than recommended (Schneider et al. 2021).

The method of defining shared nutrient requirements for a group of people who have different individual needs has its origins in the scientific nutrient requirements literature (Beaton 1995; Institute of Medicine 2000). Ethically, it follows Rawls' *maximin* principle, i.e., to maximize the welfare of the worst-off group in society, or extending to our case, to define the household diet that preferences the welfare of the nutritionally neediest member of the family (Ravallion 2016; Rawls 1971). The shared diet is most often the diet that meets *her* needs, so it is also a more gender equitable metric that can be used as a benchmark for household diets where intrahousehold allocation is not observed (Schneider et al. 2021).

---
[3] Exceeding caloric needs would lead to unhealthy diets.

Diet costs offer two evaluative functions. First, to identify populations without access to sufficient nutritious foods for better targeting of assistance. And second, to guide agricultural and food system interventions towards lower and more stable costs over time. Both the shared and individual diet cost indicators could serve these purposes, establishing an upper and lower bound respectively. The intent with these scenarios is not to describe behavior, but to use the thought experiment to demonstrate how policy decisions can be made at the household level when intrahousehold resource allocation is unobserved but differences in individual requirements are known.

The analogy with the poverty literature is illustrative. Early on, poverty lines were set at the absolute minimum income needed whereby no one would dispute that someone earning less than that amount was poor, avoiding the error of inclusion (Ravallion 1996). Similarly, "individualized" least-cost household diets, which maximize nutrient allocative efficiency within the household, can be seen as a lower bound on the affordability of an adequate diet. Inability to afford the lower bound ("individualized") diet is definitive unaffordability. Later, when poverty started to decline globally, Pritchett (2006) established an upper poverty line at the level society would consider someone no longer poor, avoiding an error of exclusion. Ability to afford the "shared" least-cost diet similarly establishes an upper bound on affordability, a level above which nobody would reasonably dispute that the household's income is sufficient to meet every member's nutrient needs.

**Data**

We combine multiple data sources to calculate monthly least-cost diets for all rural Malawian households from January 2013 to July 2017. These include: the 2013 and 2016/17 nationally representative Integrated Household Panel Surveys (IHPS) (National Statistical Office (NSO) [Malawi] and World Bank 2017), food composition data for Malawi (MAFOODS 2019), human nutrient requirements (Institute of Medicine of the National Academies 2006; Institute of Medicine of the National Academies 2011; National Academies of Sciences Engineering and Medicine 2019), and monthly market food prices from the National Statistical Office (NSO) in 25 rural markets (National Statistical Office (NSO) [Malawi] 2018).

We use the age, sex, maternity, and occupation reported in the household survey to identify individual nutrient needs for all household members. Expenditure data are used to calculate household food spending and total expenditure following domestic poverty calculation procedures (National Statistical Office (NSO) [Malawi] and World Bank Poverty and Equity Global Practice 2018; National Statistical Office (NSO) [Malawi] 2017). The price data contain monthly prices for 51 food items collected between January 2013 and July 2017 and include foods from all food groups. We match households to the market in their district or sub-district of residence (Supplementary Table A) (National Statistical Office (NSO) [Malawi] 2011; National Statistical Office (NSO) [Malawi] 2012; National Statistical Office (NSO) [Malawi] 2018). We use the rural sub-sample of the IHPS since we only have access to food price data for rural district markets (price data are collected for urban areas but not shared). Similarly, because many nutrient dense foods were added to the price monitoring list in January 2013, we do not use the first 2010 IHPS round (Kaiyatsa, Schneider and Masters 2021).

NSO purposively selected rural district markets for price monitoring. They are in the main trading towns. Despite being main centers, these towns are still relatively small; Malawi is much less urbanized than most other countries in sub-Saharan Africa (SSA) (World Bank 2021).[4] The consumer food price index computed by the NSO with these price data is considered representative of rural Malawi. The IHPS data are representative for urban and rural strata. Therefore, our results can be considered representative of rural Malawi.

Our emphasis is on the market and its ability to provide access to nutritious diets, as a metric of food system performance. Analysis at the market level provides a policy-relevant way to assess a food system, and our analysis offers the added value of ensuring the assessment is relevant for all types of individuals in realistic household settings. The prices observed in the market are for standardized items of a particular quality, so they are not directly comparable to unit costs reported by households.[5] The food price environment faced by households and their preferences are reflected in the denominator of the affordability analysis, namely the food and total expenditure calculated using reported unit costs and standard methodology to value own produced goods.[6] With two household surveys (2013 and 2016/17), we can only assess affordability in the month a household was surveyed, which is insufficient to establish seasonal patterns of affordability. Given the availability of monthly market prices between 2013 and 2017, there is, however, better scope to assess seasonality in the cost of least-cost diets (including differences by scenario). Given these different considerations (the objective of assessing market performance and the greater ability to assess seasonality in the market's ability to provide nutritious diets), we calculate the least diet costs using market prices.

Table 1 presents the characteristics of the households, household expenditure, markets, foods, and nutrients included in our sample. To establish the context for our affordability results, the median household already spends three quarters of its resources on food and lives just above the international poverty line of $1.90 per person per day in 2011 purchasing power parity (PPP) dollars (World Bank 2021).

---

[4] In 2020, only 17% of Malawians lived in urban areas, compared to 41% in SSA overall (World Bank 2021).
[5] Unit costs reflect both quality and price, and therefore differences are not necessarily attributable to different price environments (Gibson 2016; Gibson and Kim 2018; Gibson 2013).
[6] For context, Supplementary Tables B-1 (by size of difference) and B-2 (by food group) show the comparisons for items that could be compared at the district level in the same month and year, showing differences of varying magnitude (some small in practical terms) and where the market prices are more often higher but not always.

Table 1. Summary of Household Survey Data and Food Market Prices in Malawi, 2013 and 2016/17

|  | 2013 | | 2016/17 | | Overall | |
|---|---|---|---|---|---|---|
|  | Mean | (SE) | Mean | (SE) | Mean | (SE) |
| Household size | 4.76 | (0.12) | 4.98 | (0.15) | 4.90 | (0.11) |
| Number of adults (>18) | 2.32 | (0.05) | 2.52 | (0.05) | 2.44 | (0.05) |
| Number of children (≤18) | 3.51 | (0.10) | 3.52 | (0.10) | 3.51 | (0.09) |
| Food items available per month per market[a] | 39.83 | (4.94) | 39.47 | (4.59) | 39.71 | (4.82) |
| **Household Expenditure (2011 US$ PPP)** | Median | (SE) | Median | (SE) | Median | (SE) |
| Annual Food Expenditure | 2,588.47 | (113.63) | 2,292.84 | (94.08) | 2,429.85 | (91.74) |
| Per day (household) | 7.09 | (0.31) | 6.28 | (0.26) | 6.66 | (0.25) |
| Per day per capita | 1.60 | (0.10) | 1.42 | (0.06) | 1.47 | (0.05) |
| Annual Total Expenditure | 3,460.88 | (155.13) | 3,319.20 | (149.13) | 3,354.48 | (139.83) |
| Per day (household) | 9.48 | (0.43) | 9.09 | (0.41) | 9.19 | (0.38) |
| Per day per capita | 2.29 | (0.14) | 1.98 | (0.09) | 2.10 | (0.09) |
| Food Spending Share of Total Expenditures | 0.76 | (0.01) | 0.73 | (0.01) | 0.74 | (0.00) |
| **Observations**[b] |  | N |  | N |  | N |
| Households[c] |  | 1,424 |  | 1,693 |  | 3,117 |
| Individuals[d] |  | 6,995 |  | 7,907 |  | 14,902 |
| Markets[e] |  | 25 |  | 25 |  | 25 |
| Food items |  | 51 |  | 51 |  | 51 |
| Nutrients |  | 22 |  | 22 |  | 22 |

Population statistics corrected using sampling weights.
[a]Standard deviation in parentheses.
[b] Excluded: 260 infants under 6 months who are assumed to be exclusively breastfeeding, 4 rural households unable to be matched to a market.
[c] 1,081 are unique households observed at both time points, however the composition of those households changes in January 2016, so these are best thought of as two repeated cross sections of households and individuals.
[d] Excludes individuals who reported eating no meals in the household in the prior week, allowing diet cost to be compared to reported food consumption expenditure.
[e] List of markets and districts provided in Supplementary Table A.

We identify the nutrient content of foods using the Malawi Food Composition Table (MAFOODS 2019), the USDA National Nutrient Database for Standard Reference (where necessary) (USDA 2018) and using conversion factors provided by the NSO.[7] To perform seasonality analysis at the food group level, we classify foods into the groups used for dietary diversity indicators (WHO 2008; FAO and FHI 360 2016; Kennedy, Ballard and Dop 2010; Ministry of Health (MOH) [Malawi] 2017). Supplementary Tables C and D present the food item sources of each nutrient and the food items within each food group. Prices are monitored for most foods that households choose to consume (Supplementary Table E).

The least-cost diets can only select from the menu of 51 items included in the price dataset and where there is a price observation. To explore concerns about how missing food price data (a common feature of consumer price index datasets) could affect the results, Figure 1 shows the pattern of price records across all 51 food items by month. The darkest green cells indicate a price was observed for that item in all markets. Bias may result, for example, if prices erroneously reflect unavailability of a food item and a solution to the least-cost diet optimization proves infeasible or the calculated cost of the least-cost diet is higher than it would have been if the food item could have been considered in the optimization problem (i.e., the price was present). Similarly, there may be bias if the price of a food item is recorded, but the food is in fact not available in the market.

Note, however, that linear optimization is a data envelopment technique. So, any missing or erroneous data would only affect the results if that item *would have otherwise been selected into the diet* were the true data value known. There are multiple food sources for all essential nutrients, and all food items contain multiple nutrients, meaning many food combinations could meet the minimum requirements, reducing the likelihood that a single erroneously missing price would lead to infeasible results or substantially increase the cost of the least-cost diet. There might also not be any solution to the linear optimization if there is no combination of foods that could meet the minimum requirements while also staying under all the upper limits and including the exact amount of energy required (even if the food for which the price is missing were truly available). In practice, no single food item is likely so optimal in nutrient density for all individuals and households, so that the likelihood that its missing price would be binding in all cases is small.

Turning to the market price data (Figure 1), we note that at least one form of maize, the staple food, is reportedly available in all markets in all months. The items for which market prices are most commonly unreported are perishable fruits and vegetables whose production is highly seasonal, suggesting that missing prices reflect periods when those foods are not being sold in the surveyed markets, i.e., they are likely to reflect real unavailability. For several frequently missing items, there are also substitutes in the food list (e.g., cooking oil and cooking oil refill), and for each month no food group has all items missing simultaneously. Some items may never be present in a location (e.g., fresh *chambo* fish), but other nutritional substitutes (e.g., other fish) are present.

In follow up work Kaiyatsa et al. (2021) further employed a special market survey in the same rural Malawian markets to investigate whether missing prices in the 2013-2017 price data are mainly related to (seasonal) unavailability of the food (thus reflecting market reality) or whether they were more likely due to measurement error (foods available, but prices not recorded). Comparing the 2013 – 2017 price data to the survey results, 17% of the price observations were missing when reportedly usually available in that market and month ("discordant missing") and 3.9% had a price observation when reported usually unavailable ("discordant present") (likely due to *ex post* imputation). The discordant missing prices were not meaningfully explained, however, by food item, time, market, or their interactions, suggesting that substitution possibilities are likely and that there is little systematic bias in the least-cost diet calculations by market or month. They further estimate that in markets with higher than average discordant missing (present) prices, the least-cost nutrient-adequate diet for an individual non-pregnant, non-lactating woman of reproductive age is biased 3.9% (7.6%) upward (downward). These results lend confidence that even where missing prices are due to measurement error, the magnitude of bias introduced into the cost estimate is reasonably small.

---

[7] Specific information regarding food item composition matching records are available in the replication data files and we also point readers to the MAFOODS data tables (MAFOODS 2019). Further detail on the procedures to convert foods into their nutrient components available in (Schneider 2021; Schneider et al. 2021)

Figure 1. Frequency of observed prices by item, month, season, and food group

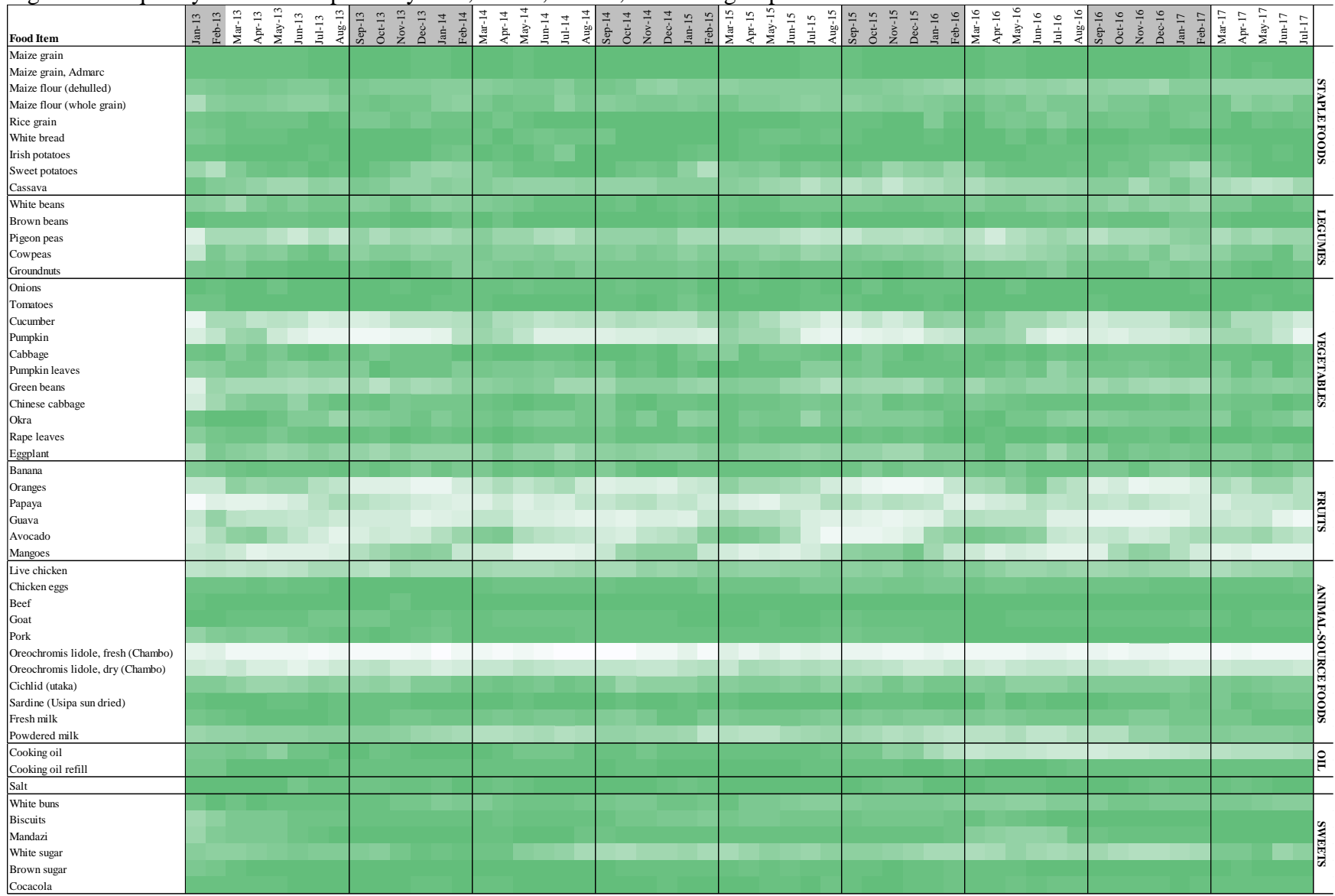

**Figure 1:** Frequency of observed prices by item, month, season, and food group. Illustrates the number of markets where a price was observed price for each item in each month is shown by the darkness of the tile. Darkest (lightest) item-months indicate the price was observed for that item in that month in the highest (fewest) number of markets.

## Methods

*Individual Nutrient Requirements*

Biological nutrient requirements for individuals are specified by age, sex, maternity status, and physical activity level by the Institute of Medicine (Institute of Medicine of the National Academies 2006; Institute of Medicine of the National Academies 2011; National Academies of Sciences Engineering and Medicine 2019). These requirements dictate lower and upper bounds for all nutrients that must be consumed through food. We include constraints for energy, macronutrients (carbohydrates, protein, fat), and all the micronutrients (vitamins and minerals) where there is sufficient scientific evidence to set an average minimum requirement at the population level (vitamins A, C, E, B6, and B12, thiamin, riboflavin, niacin, folate, calcium, copper, iron, magnesium, phosphorus, selenium, zinc, and sodium). Twelve nutrients have an upper and lower bound, seven have only a lower bound and no upper limit, and retinol and sodium have only an upper bound (National Academies of Sciences Engineering and Medicine 2019; Institute of Medicine of the National Academies 2006; Institute of Medicine of the National Academies 2011). We calculate energy needs based on reference heights and weights for each age and sex, account for breastfeeding, and use occupation to estimate physical activity level (for further detail see Schneider and Herforth 2020; Schneider 2021; Schneider et al. 2021). We refer to these scientifically defined nutrient requirements as the "individual" requirements, and they are the constraints in solving the lower bound (individualized diets) least-cost diet problem.

*Household Nutrient Requirements*

We follow Schneider et al. (2021) to establish the shared household nutrient requirements. Formally, for each household (h) the requirement considers each individual household member's (i) requirement for each nutrient (j) and their energy needs ($E$), using the most restrictive nutrient density requirements for any member:

(1) $\quad Lower_{hj} = \sum_i E_i * max_i \{MinimumNeed_{j,i}/E_i\}, \; j = 1, \dots, 19$
(2) $\quad Upper_{hj} = \sum_i E_i * min_i \{MaximumTolerance_{j,i}/E_i\}, \; j = 1, \dots, 13$
(3) $\quad HHE_h = \sum_i E_{ih}$

Equations (1-3) are used to compute shared requirements for household members four years and above. To this we add individual requirements for children six months through three years old to arrive at the household total for which the least-cost diet problem is solved. More information on the nutrient requirements, requirement tables, and comparison of the individual and shared nutrient requirements can be found in Schneider et al. (2021) and Schneider (2021).

*Cost of Nutrient Adequacy (CoNA) Index Construction*

Using linear programming, we attempt to identify a diet that meets all the specified nutrient requirements at the lowest total cost. Formally, the linear optimization model (solved using the R package "lpSolve" by Buttrey (2005)) minimizes total cost over all foods (f) within upper and lower bounds for all nutrients (j) and meets the specified energy budget ($E$). Adding data on price ($p_f$) for each food item (f) and its nutrient contents ($a_{fj}$) yields:

(4) $\quad CoNA: minimize\ C = \sum_f p_f * q_f$

Subject to:
$\sum_i a_{fj} * q_f \geq Lower_j,\ j = 1, \dots, 19$
$\sum_i a_{fj} * q_f \leq Upper_j,\ j = 1, \dots, 13$
$\sum_i a_{fe} * q_f = E$
$q_1 \geq 0, q_2 \geq 0, \dots q_i \geq 0,$ for all foods $i = 1, \dots 51$

Equation (4) is solved every month for everyone (with individualized requirements) and for each household (with shared requirements, i.e., replacing $Lower_j$, $Upper_j$ and $E$ by equations (1), (2) and (3) respectively), using the foods and prices in the market of the household's district of residence. The household's "individualized" least-cost diet (lower bound) is obtained as the sum of its individual least-cost diets.

We compute least-cost diets at the monthly level. Nutrient requirements corresponding to the observed demographics in 2013 are used to solve the diet cost problem from 2013 through 2015, and then the household composition and corresponding nutrient requirements observed in 2016/17 are used to solve the diet cost problem from January 2016 forward. Food expenditures were collected for the previous seven days and therefore reflect the consumption of those who ate in the household in that period (Fiedler and Mwangi 2016). We scale the nutrient requirements for any partial meal-taking to accurately draw comparisons with observed food spending. For every household, 55 *CoNA* indices (36 observations from 2013-2015 and 19 observations from January 2016 to July 2017) are obtained per scenario.

We focus on two primary results from the linear modeling: feasibility of a solution and cost. We use the binary outcome of a solution or no solution to summarize the extent to which there is a feasible diet in each month. Under the individualized diets scenario, we consider the household to have a least-cost diet solution only if there is a solution for all members. As discussed above, confidence that no solution to the least-cost diet problem reflects realistic infeasibility rests on three observations. First, patterns observed in Figure 1 show that most missing prices are for perishable items, reasonably understood to be seasonally available, and there are consistently some items from each food group (closest nutritional substitutes) with price observations. Second, we refer to the study by Kaiyatsa et al. (2021), which found the pattern of missing prices was not explained by the food item, time, or market. Third, the fact that individual diets are feasible in most markets and months provides further confidence that if shared diets are not feasible, it is less likely to be resulting from underlying missing price data than due to the nutrient requirements. Thus, we conclude that no solutions to the least-cost diet problem corresponds to realistic infeasibility in that market/month, given the available items and the nutrient requirement constraints applied.

If the model finds a solution, we convert all costs into 2011 US$ PPP using the Denton method to smooth annual conversion factors over months (World Bank 2015; Denton 1971; International Monetary Fund 2018).

*Seasonality*

Food availability and prices vary across and within years. However, since the linear programming model will substitute among foods given availability and prices, seasonality in the prices of individual foods does not necessarily carry over to the same extent into *CoNA* indices.

Intra-annual fluctuations can be regular and stochastic. Here the focus is on regular intra-annual fluctuations, or seasonality, of the *CoNA* index and relative to that of maize and main food groups. A standard indicator to measure seasonality is the seasonal gap, the ratio between the seasonal peak and trough. For food prices in low-income country settings characterized by one growing season per year, these are most observed just before and just after the harvest, respectively (Gilbert et al. 2017).

Linear detrended seasonal dummy and moving average deviation models are often used to estimate seasonality. They also have some limitations, especially the potential introduction of upward bias with short time series such as our dataset (see Gilbert et al. (2017) for a comprehensive discussion). Trigonometric models have been shown to address some of these limitations because they are parsimonious in the number of parameters needed for estimation and are less prone to biased gap estimation, especially for shorter time series (Ray et al. 2001; Kaminski et al. 2016; Bai et al. 2020; Kotu et al. 2019; Wassie, Kusakari and Sumimoto 2019). Following Gilbert et al. (2017) and comparing both methods, we also found the trigonometric method to be preferred by model fit statistics (Supplementary Table F) for the food price dataset. For the *CoNA* results, the stochastic trend model was preferred.

The stochastic trend dummy model we use allows for gaps of (k) months prior to the observation (a) in time *(y,m)* as follows:

(5) $\quad \Delta_k C_{hym} = C_{hym} - C_{hy,m-k-1} = k\gamma + \sum_{a=1}^{k-1} \delta_{m-a}(s_{m-a}) + w_{hym}$

Where *C* is the log cost in nominal terms. The cost in nominal terms is used since food expenditure comprises a large proportion of budget shares. Therefore, deflation is sensitive to food prices and their use in seasonality analysis may understate the extent of seasonality (Gilbert et al. 2017). And *k* is the number of month gaps between the current month observation and the preceding observation. In our case, gaps are the household-months with no solution to the linear programming problem. Where there are gaps, differences are calculated as the difference between a diet cost observation and the most recent preceding observation. The seasonal differenced dummies are then defined as:

(6) $\quad s_{m-a} = \begin{cases} 1 & a = m \\ -1 & k = 0 \\ -1 - k & k > 0 \\ 0 & \text{otherwise} \end{cases}$

And seasonal factors are calculated by demeaning the coefficients.

When we analyze seasonality, we are limited to household-months with a feasible solution to the least-cost diet, which introduces selection bias. To estimate the magnitude of this potential bias, we consider no solution to the least-cost diet problem as akin to having very high (possibly infinite) cost.[8] We estimate the seasonality model first with these months recorded as missing, and then with costs for those months imputed as the highest cost observed over all markets and households in that same month and year. Repeating the seasonality analysis using these imputed data allows us to estimate a lower bound on the magnitude of the true seasonality in the diet cost.

---

[8] This assumes that a no-solution represents infeasibility of a solution in the market (and not because of erroneously missing market prices), which we have shown above to be realistic in the context of our data.

To further explore the role of least-cost diet availability in our least-cost diet seasonality estimates, we examine the prevalence of a feasible diet, conditional on the month and by scenario, modeled with a linear probability model as follows:

(7) $\quad A_{hym} = \gamma + month_m + market_l + \mu_{hym}$

Where *A* is a binary indicator of diet feasibility for household (h) in time (y,m), and the probability of feasibility is estimated for each scenario with indicator variables for each calendar month (m) and market location (l). The seasonal factors on feasibility are calculated following equation (6) by demeaning the coefficients. These are subsequently compared to the seasonality in costs.

We model the seasonality in underlying food prices to compare with the diet cost seasonality using the trigonometric model. We calculate the difference in logged price for each food item (in nominal terms), allowing gaps where no price was observed, as follows:

(8) $\quad \Delta P_{hym} = \gamma + \alpha \Delta \cos\left(\frac{m\pi}{6}\right) + \beta \Delta \sin\left(\frac{m\pi}{6}\right) + \mu_{hym}$

Where *P* is the logged price per kilogram edible portion in nominal terms, for household (h) in year (y) and month (m). The seasonal factors can be computed as follows:

(9) $\quad S_m = \lambda \cos\left(\frac{m\pi}{6} - \omega\right)$
$\quad\quad$ where $\lambda = \sqrt{\alpha^2 + \beta^2}$ and $\omega = \tan^{-1}\left(\frac{\alpha}{\beta}\right)$

We then regress the difference in price on the monthly indicator variables, pooling items in each food group. Food groups classify items into nutritionally relevant categories, those that might be substitutes in the linear programming. Greater seasonality would be expected with short harvest periods, perishability, and groups with few items. Given the importance of maize in Malawian diets, we present the same seasonality analysis for maize prices, separating maize grain in regular retail markets and maize grain sold by the parastatal Agricultural Development and Marketing Corporation (Admarc). This also facilitates comparison with other food price seasonality results reported in the literature for Malawi.

*Affordability*

Affordability is measured by relative expenditure ratios comparing the daily cost for the whole household, per scenario, to daily food or total expenditure, in nominal terms. Expenditure accounts for the contribution of own produced goods in household consumption. The premium for the shared diet is expressed through the ratio of shared to individualized diets daily cost, in nominal terms. We only assess the affordability for households with a solution to the least-cost diet under both scenarios in their month of survey. To draw conclusions about the whole sample we use only the observation for each household in the month it was surveyed. If the diet was feasible, we then also assess affordability. We define a household as having access to the diet if there is a feasible solution *and* the diet cost is affordable (relative to food and total expenditure, respectively) *in the month of survey*.

## Results

*Feasibility & Cost*

Figure 2 shows that, as expected, individualized diets are consistently more feasible than shared diets (90% of household-months relative to 60% respectively).

Figure 2. Feasibility of household nutritious diet, by month and scenario

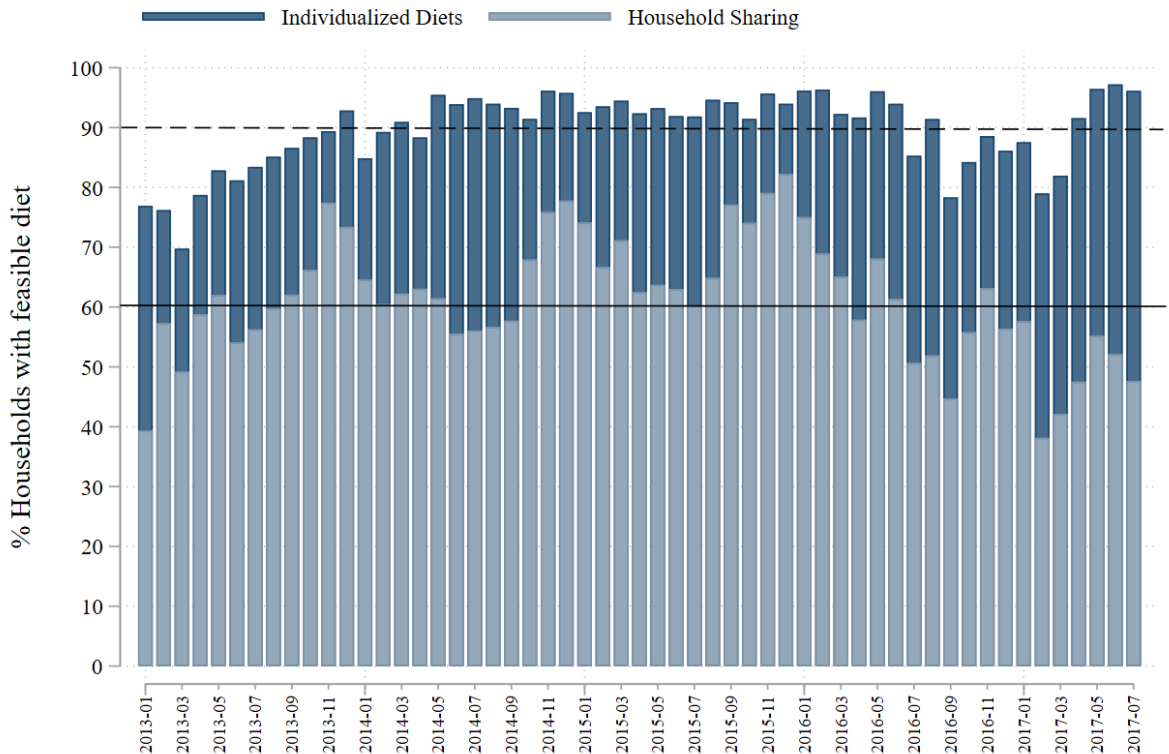

Notes: Population statistics corrected using sampling weights. Under the individualized diets scenario a household is defined to have a feasible diet in the market if all its members have a feasible diet. Solid line shows mean feasibility for shared diets (60.2%). Dashed line shows mean feasibility for individualized diets (89.7%).

**Figure 2:** Feasibility of household nutritious diet, by month and scenario. Dark (light) bars indicate the percentage of households with a feasible diet in that month under the individualized (shared) diets scenario, with dashed (solid) line indicating mean feasibility.

If an adequate diet is not feasible at the individual level, it mainly concerns children six months through three years, breastfeeding women, and older adults (70+ years) (see Supplementary Table G). These age-sex groups have a greater need for nutrient density, and for young children they are also more sensitive to toxicity and therefore have a lower tolerance at the upper limit for certain nutrients, tightening the constraints on the optimization problem relative to other age and sex groups (Schneider et al. 2021). Since the equation is estimated with the same items and prices for all members of a household, households with these types of members are therefore less likely to have a feasible diet.

When the individualized diet is feasible for a household, but the shared diet is not, it means that the foods are there, their price and nutrient composition are observed, and that a combination can satisfy individuals' nutrition needs but not the more nutrient-dense shared requirements. The infeasibility derives from higher nutrient-density required and/or tighter upper limits required for the shared diet. More diverse and larger households are more likely to have members with different nutrient requirements that further restrict the nutrient density of the diet required to meet every member's needs. This is also borne out empirically; the feasibility of a shared diet is systematically related to both household composition and household size. Thus, as a food system metric it is more likely to identify lack of access to an adequate diet for households with particularly vulnerable members such as children six months through three years, breastfeeding women, and older adults (70+ years) than other household-level approaches (Schneider, 2022).

Furthermore, and consistent with the notion of lower and upper cost bounds, when both individualized and shared diets are feasible for a household, individualized diets also tend to cost 21% less ($1.79 per capita per day on average relative to $2.26 respectively) in US dollars at 2011 PPP prices; Figure 3). These costs compare with the international extreme poverty line of $1.90/day/person (US$ 2011 in PPP terms), indicating that nutritionally adequate diets are not available for sizeable shares of Malawi's population, a point we return to below. [9] Visual inspection of Figures 2 and 3 further hints at the potential existence of seasonal patterns in the feasibility and cost of nutritionally adequate diets.

---

[9] These compare with an unweighted population average individual cost of a nutrient-adequate diet in Malawi calculated by (Herforth et al. 2020) of $1.29/person/day (2011 US$ PPP), which were calculated using different food price and nutrient composition data for global comparability. That paper used the food list and prices from the World Bank's International Comparison Project and used USDA food composition data (Herforth et al. 2020).

Figure 3. Cost of household nutritious diet, by scenario

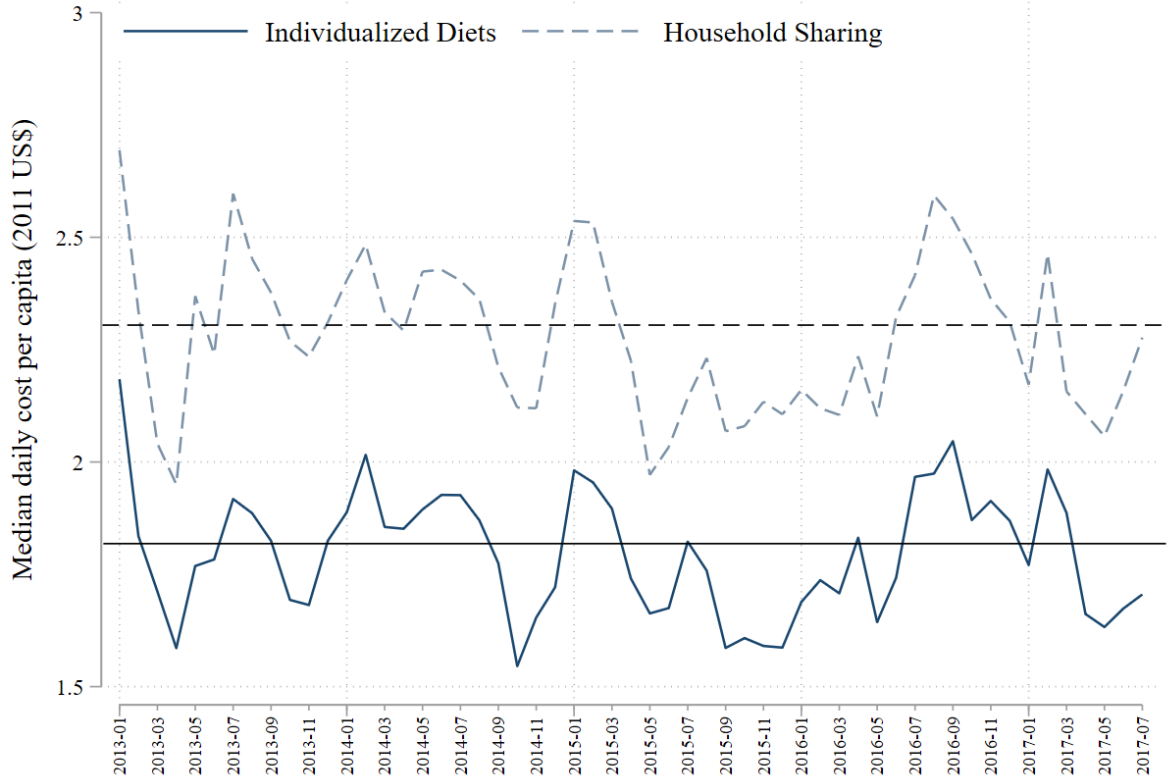

Notes: Population statistics corrected using sampling weights. Solid horizontal line shows mean cost of $1.79/capita/day (2011 US$) for individualized diets. Dashed horizonal line shows mean cost of $2.26/capita/day for shared diets (2011 US$).

**Figure 3:** Cost of household nutritious diet, by scenario. Illustrates monthly variation in the cost of the individualized diet (solid line) and shared diet (dashed line). Horizontal lines indicate the mean cost for each diet scenario.

*Seasonality*

Table 2 presents the results of the extent of seasonal fluctuation in nutritionally adequate diets more systematically. Column 1 shows the percent of households with a feasible diet by month, obtained by estimating equation (7). Column 2 shows the monthly seasonal factors when least-cost diets are feasible, obtained by calculating equations (5) and (6). Column 3 reports the monthly seasonal factors with the cost of infeasible diets imputed, obtained by estimating equations (5) and (6) using imputed cost values for the diet cost. The seasonal factors present the percentage point difference between the monthly conditional mean cost/feasibility and the grand mean; the seasonal gap is the percentage point difference between the highest and lowest seasonal factor.

Table 2. Seasonal Variation in Diet Feasibility and Cost, 2013–2017

| | (1) Household-Months With Feasible Diet | | (2) Seasonal Variation in Diet Cost *if feasible* | | (3) Seasonal Variation in Diet Cost *imputing infeasible* | |
|---|---|---|---|---|---|---|
| | *Estimated Percent Feasible*[a] | | *Seasonal Factors*[b] | | *Seasonal Factors*[b] | |
| | Individualized Diets | Household Sharing | Individualized Diets | Household Sharing | Individualized Diets | Household Sharing |
| January | 91.2 | 58.4 | -1.7 | 0.6 | 4.2 | 42.2 |
| February | 89.7 | 52.5 | 3.3 | 3.1 | -3.6 | 31.3 |
| March | 88.8 | 52.3 | -0.1 | 0.7 | -10.9 | -8.0 |
| April | 92.0 | 52.0 | 0.9 | 2.3 | -13.4 | -9.5 |
| May | 96.2 | 57.3 | 7.2 | 4.3 | -0.4 | -11.7 |
| June | 95.2 | 52.4 | 4.9 | 2.5 | 18.5 | 21.8 |
| July | 93.1 | 48.3 | -1.0 | -1.8 | 8.3 | -18.4 |
| August | 93.9 | 52.6 | -5.4 | -4.4 | -4.9 | -51.5 |
| September | 89.2 | 53.3 | -6.6 | -5.8 | 5.1 | -38.6 |
| October | 90.7 | 59.7 | -3.7 | -2.1 | -11.7 | -33.4 |
| November | 94.4 | 67.5 | 2.6 | 0.8 | -1.6 | 11.6 |
| December | 93.8 | 65.3 | -0.3 | -0.2 | 10.5 | 64.1 |
| ***Seasonal Gap*** | 7.4 | 19.2 | 13.8 | 10.1 | 31.9 | 115.7 |
| Mean Availability[c] | 89.7 | 60.2 | | | | |
| (% Household-Months) | (0.9) | (1.6) | | | | |
| Median Cost, per capita per day[c] | | | 1.79 | 2.26 | | |
| (2011 US$ PPP) | | | (0.03) | (0.04) | | |

[a] Calculated as in equation (8), interpreted as the percent of households with a feasible diet on average each month. The seasonal gap in feasibility is the percentage point difference between the most feasible and the least feasible month.
[b] Seasonal factors of diet cost calculated as in equation (6) interpreted as the percentage point difference in average cost in that month relative to the average over all months of the year.
[c] Standard error in parentheses.

The following insights emerge. First, the seasonal gap in feasibility (Column 1) is 7% and 19%, for individualized and shared diets, respectively. The seasonal gap in costs (Column 2) is 14% and 10%, respectively (when the least-cost diets are feasible). This shows non-negligible seasonality both in the feasibility and the cost (when feasible) of nutrient-adequate diets in Malawi.

Second, shared diets display substantially more seasonality than individualized diets, which mainly follows from their greater infeasibility. To consider this, infeasible diets are imputed by the highest observed cost by month and year (per scenario) (Table 2, Column 3). This arguably still puts a lower bound on the extent of seasonality but aims to address the selection bias introduced into the estimation by infeasibility. Doing so yields a seasonal gap of 31 percentage points for individualized diets, which is more than twice the gap observed when looking at feasibility or costs only when feasible. For shared diets, the seasonal gap increases to 115 percentage points, an 11-fold greater gap than that observed only when the diets are feasible. It indicates that shared diets display substantially more seasonality than individualized diets and that this is mainly due to the greater infeasibility of shared diets across seasons.

Third, turning to the pattern of seasonality, the shared diet is more feasible during October-January, peaking in December (Table 2, Column 1), even though these months are typically considered the lean season. They are most costly during April-June, the harvest months (Table 2, Column 2). One potential explanation for the former is the greater availability of nutrient dense animal-source foods during those months for cultural reasons; many Malawians consume meat during the holidays (particularly Christmas). More fish is also available in those months prior to the fishing ban associated with the spawning season (FAO 2005; Gelli et al. 2020; Gilbert et al. 2017). Schneider (2022) also demonstrates that the adequate shared diets contain a larger share of legumes so that all nutrient requirements can be attained without exceeding copper upper bounds. Legumes are harvested in the months prior, during, and after the maize harvest (~March through August depending on the specific type) (Freer et al. 2018). Though they are a more expensive source of calories than staples, they enter the diet to satisfy the micronutrient constraints (specifically sufficient selenium without exceeding the copper upper limit) thereby providing a greater share of calories than would be most cost efficient if other sources of micronutrients were available. However, when imputing the cost of infeasible diets (Table 2, Column 3), the seasonality pattern of nutritiously adequate diets is largely consistent with the commonly observed harvest/hunger cycle (lower dietary costs following the harvest months (with the exception of June) and higher dietary costs during the lean months), underscoring the importance of the market availability of sufficiently diverse foods throughout the year.

To further understand the extent to which the findings on seasonality in dietary adequacy reflect seasonality in food prices, we compare seasonality in the diet cost to seasonality in food item prices, by food group (Figure 4). First, vitamin A-rich vegetables (pumpkin) display the largest seasonal gap (57.8%), followed by maize, dark green leafy vegetables, roots and tubers and fruits (vitamin A-rich), which all have a seasonal gap of 20-25%. Seasonality is lowest for milk, eggs, fish, and meat, as to be expected for items that can be produced year-round. Overall, the findings are similar to those reported in the literature, for Malawi as well as neighboring countries (Bai et al. 2020).

The high seasonality in vitamin A-rich vegetables is likely driven by having price data for only one item (pumpkin) in this food group so the food group follows its harvest pattern. Furthermore, price seasonality in the cereals food group (maize as well as rice, cassava, Irish and sweet potatoes, bread) is only about half (12.1%) that of maize alone (21.0% for retail market, 24.5% for Admarc maize grain). This is consistent with findings by Manda (2010) and Gilbert et al. (2017) that seasonality in maize prices is about twice that of rice. This suggests that consumers could smooth consumption by switching away from maize to other staples when prices are high. At the same time, seasonality in dietary costs is larger than that of cereals (for both individualized and shared diets), suggesting that seasonality in the cost of nutritionally adequate diets is mainly driven by the availability and price seasonality of other food groups. Consistently, seasonal peaks in dietary costs (calculated for observations where diets are feasible) are observed when maize prices are at their trough, i.e., during the maize harvest season.

Figure 4. Seasonal gap in food group prices, by food group

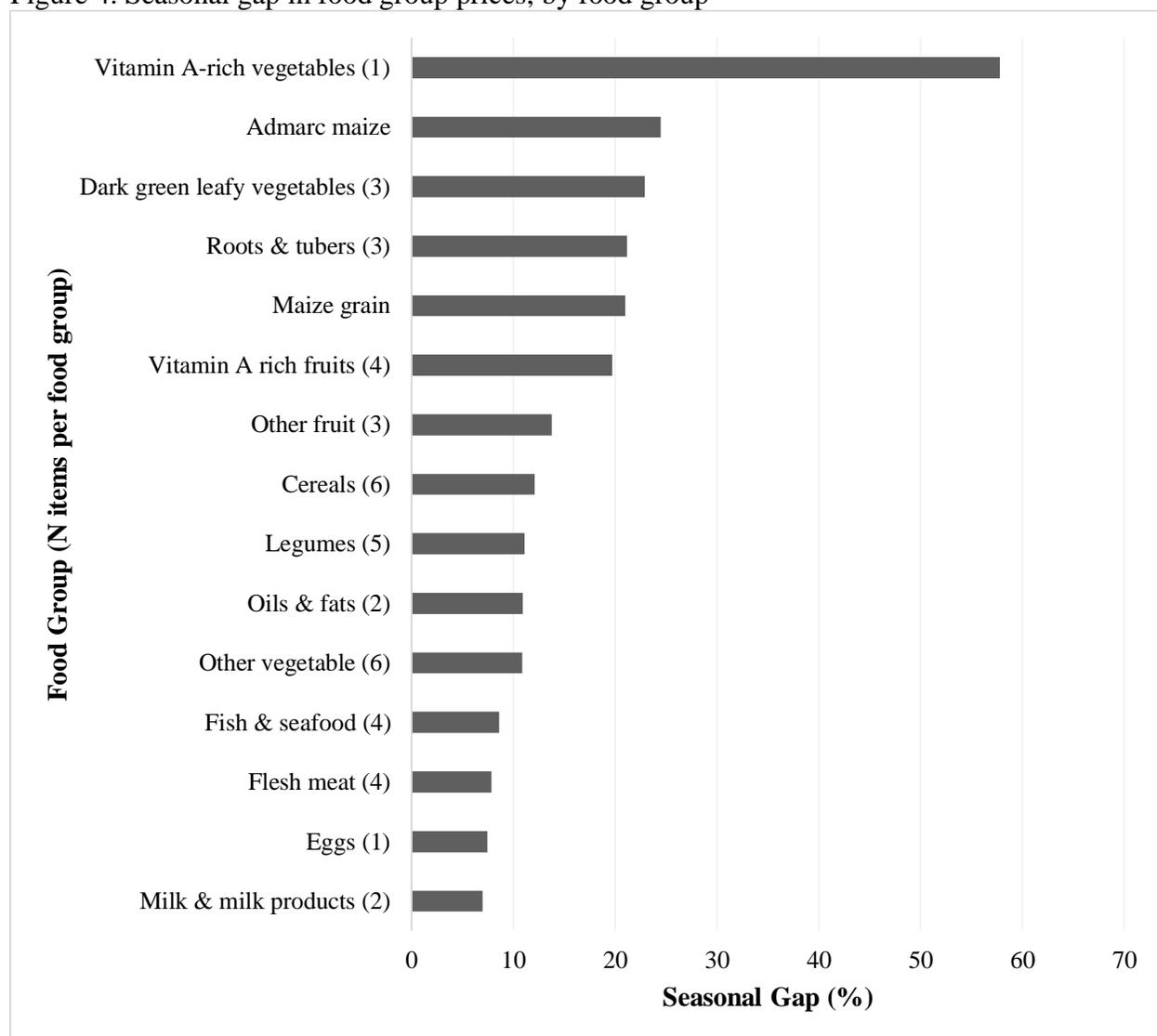

Notes: Heteroskedasticity robust standard errors clustered at the market level. Pooled trigonometric regression estimated by food group. Cereals includes maize grain, Admarc maize grain, maize flour dehulled, maize flour whole grain, rice, white bread. The single item in the vitamin A-rich vegetables food group is pumpkin. Although orange-fleshed sweet potato has become widely disseminated in Malawi in recent years (Low et al. 2017; Low and Thiele 2020), the NSO collects prices only for white sweet potatoes and Irish potatoes. Sweets and condiments excluded from seasonality analysis. The seven items not included in this figure are: Coca-Cola, biscuits, scones/buns, *mandazi* (fried dough), white sugar, brown sugar, salt. For the "Other fruit" group, the BIC was equivalent for fixed effects dummy and trigonometric specifications.

**Figure 4:** Seasonal gap in food group prices, by food group. Shows the average gap over all items in the food group, with number of food items per food group in parentheses.

Overall, the findings on cost seasonality suggest that the degree of seasonality in the cost of nutrient-adequate diets is 1) non-negligible, 2) at least as large as that of individual food groups (larger than that of cereals), and 3) larger for shared than for individualized diets. Further, the seasonality in cost of the diet is mainly driven by feasibility and non-maize food groups.

*Affordability*

Thus far we have discussed the extent and seasonality of dietary costs on their own. Table 3 assesses the costs of these diets relative to the food and total expenditure for households who have a least-cost diet solution *in their month of survey*. For individualized diets, the median cost ($1.83/person/day) is 11% higher than current per capita food expenditure and equivalent to 78% of monthly total expenditure. For shared diets, the median cost ($2.31/person/day) is 35% higher than current food spending and 92% of monthly total expenditure. Comparing shared with individualized dietary costs, the premium for household sharing is 33%. This is possible only when both the shared and individualized diets are feasible and present a lower bound on the premium.

Table 3. Nutritionally Adequate Diet Feasibility, Cost, and Affordability in Month of Survey

|  | 2013 – 2015 | | 2016 – 2017 | | Overall | |
| --- | --- | --- | --- | --- | --- | --- |
|  | Median | (SE) | Median | (SE) | Median | (SE) |
| **Lower Bound: Individualized Diets** | | | | | | |
| Households with a feasible diet in month of survey (%) | 83.71 | (3.02) | 88.45 | (2.39) | 86.73 | (2.12) |
| Cost per day (household) | 7.51 | (0.26) | 8.31 | (0.37) | 7.96 | (0.29) |
| Per capita | 1.85 | (0.07) | 1.82 | (0.08) | 1.83 | (0.05) |
| Per 1,000 kcal | 0.95 | (0.04) | 0.94 | (0.04) | 0.94 | (0.03) |
| Cost/Food Expenditure | 1.01 | (0.05) | 1.19 | (0.07) | 1.11 | (0.06) |
| Cost/Total Expenditure | 0.72 | (0.04) | 0.81 | (0.05) | 0.78 | (0.04) |
| N households with a solution in month of survey | 1,125 | | 1,451 | | 2,576 | |
| N households with no solution any month in year of survey | 37 | | 52 | | 89 | |
| **Upper Bound: Shared Diet** | | | | | | |
| Households with feasible diet in month of survey (%) | 58.98 | (3.71) | 54.50 | (3.25) | 56.13 | (2.58) |
| Cost per day (household) | 9.49 | (0.47) | 9.13 | (0.43) | 9.24 | (0.36) |
| Per capita | 2.39 | (0.10) | 2.25 | (0.08) | 2.31 | (0.06) |
| Per 1,000 kcal | 1.26 | (0.05) | 1.20 | (0.04) | 1.21 | (0.03) |
| Cost/Food Expenditure | 1.24 | (0.09) | 1.40 | (0.09) | 1.35 | (0.06) |
| Cost/Total Expenditure | 0.88 | (0.06) | 0.95 | (0.07) | 0.92 | (0.04) |
| N households with a solution in month of survey | 792 | | 956 | | 1,748 | |
| N households with no solution any month in year of survey | 187 | | 223 | | 410 | |
| **Scenario Comparison (annualized)** | | | | | | |
| Shared Cost/Individualized Diets Cost | 1.34 | (0.02) | 1.31 | (0.02) | 1.33 | (0.02) |
| N Households with solution under both scenarios in month of survey | 727 | | 897 | | 1,624 | |
| Households (total N) | 1,424 | | 1,693 | | 3,117 | |

Notes: Population statistics corrected using sampling weights, standard errors clustered at the enumeration area level. This table reflects the data for each household in the survey month, where the median is $1.83/person/day. This differs from Table 2 which includes all data for all households and all months and finds the median to be $1.79/person/day.

If we also classify households for whom the diet is infeasible as not being able to access the diet, the individualized diet is infeasible or costs more (in the market) than total expenditure for 44% of rural Malawian households.[10] For an additional 18%, the individualized (lower bound) diet is feasible, but unaffordable without increasing current food expenditure (though technically affordable within total expenditure). In total, almost two in three rural Malawians (62%) cannot access any adequate diet in the market because even the lower cost individualized diet is infeasible, costs more than they have at all (total expenditure), or costs more than they currently choose to allocate to food.

Even fewer households have access to the shared diet. For 69.5% of rural households, the diet is infeasible or costs more than total expenditure. For an additional 10.5%, the diet is feasible but costs more than current food spending, though less than total expenditure. Only 20% of rural Malawian households face a feasible shared diet that stays within their current food budget. In between those who cannot afford the individualized diets without increasing food spending (our lower bound on the cost of nutrient-adequate diets) and those who cannot afford the shared diets without increasing food spending (our upper bound on the cost of nutrient-adequate diets), we identify 18% of the population who could afford to meet the family's nutrient needs if carefully allocating foods to different household members according to their needs to achieve nutrition goals.

**Conclusions**

In this paper, we developed two methodologies to estimate least-cost diets for whole households and used empirical analysis of Malawi's markets to illustrate the implications of taking seriously imperfect targeting of intrahousehold food and nutrient allocation given differing individual needs. The two methods produce a lower and upper bound on the cost of a nutrient-adequate diet for a whole household in the market. They essentially reflect perfect nutrient targeting through individualized diet cost optimization (lower bound) and a Rawlsian approach to household diet cost optimization adopting the nutrient density of the neediest member and assuming full meal sharing according to energy needs (upper bound).

In the case of Malawi, the higher diet quality demanded for shared meals is substantially less likely to be feasible in the market than if households were to pursue perfectly individualized diet strategies (60% versus 90% of all market-months studied). Shared diets also cost 33% more, on average, when feasible. Taking current per capita food spending as benchmark, at least 62% of rural Malawian households lack access to a nutrient-adequate diet, because the individualized diet is infeasible, or its cost exceeds current food expenditure. Clearly, increasing food budgets without increasing incomes would be near-impossible for many (recall that households already spend an average 74% of their expenditures on food). Nonetheless, the situation is even more dire when accounting for meal sharing—with 80% of households not able to afford the shared diet within current food budgets. With the latter result driven by greater sensitivity to the nutritional needs of women and girls (as well as the very young and very old), shared diets also bring important gender and equity dimensions to the nutrient adequacy assessment of current food systems.

---

[10] For clarity, we emphasize this summary statistic reflects one observation per household per survey round and the cost/feasibility result in the month in which the household was surveyed, concurring with the timing that all other data were collected, regardless of the feasibility of the diet. Thus, (with survey weighting) we can accurately draw population-level conclusions.

Despite the potential to substitute across foods to obtain the required nutrients, access to nutrient-adequate diets still varies considerably across seasons. The seasonal gap in the cost of the individualized (lower bound) diet is estimated at 14% when the diet is feasible and at least 30% when considering infeasible diets to have the highest observed cost. This result provides an estimate for the unavoidable extent of seasonality in diet costs, or the greatest possible smoothing of diet costs under current market conditions in Malawi. For shared diets, the seasonal gap has an estimated lower bound of 116%.

Thus, seasonality in diet costs in Malawi's current food system remains substantial and cannot be ignored. It is higher than price seasonality among cereals (12%) and at least as high as price seasonality among most food groups. It is mostly driven by the infeasibility of nutrient-adequate diets during certain months given the available foods in the market, which points to the importance of appropriate food policies and investments (including irrigation) to make nutritious foods available and affordable, year-round. Given the currently available foods in rural markets in Malawi, guiding consumers to pursue more individualized diet strategies could further help to smooth access to nutritious diets throughout the year, so that, for instance, legumes and vegetables could be disproportionately eaten by women, young children, and the elderly as opposed to men and boys who need a smaller amount of these foods to meet their nutrient requirements beyond calories and can meet their higher energy needs from cheaper (staple) sources (Schneider 2021).

Two key policy recommendations emerge, for Malawi, and for assessing the dietary adequacy of food systems more broadly.

First, substantial investments in agriculture and rural areas will be needed in Malawi to raise rural incomes and make a wider range of foods (besides maize) and nutritionally adequate diets more available and affordable throughout the year. While this is the first order food and nutrition policy issue for Malawi, the calculation of the bounded range on dietary adequacy further shows that additional strategies are also warranted. Given the currently available foods in its rural markets, behavior change communication and other nutrition education interventions are advisable to encourage households to target nutrient dense foods to the neediest members, including to reduce seasonality in dietary adequacy (Sanghvi et al. 2013; Schneider et al. 2021; Ruel and Alderman 2013; Gillespie and van den Bold 2017).

Second, least-cost diets can also be used to identify households who cannot afford a nutritionally adequate diet and calculate the amount of public assistance needed (Carlson et al. 2007). Individualized diets provide a starting point but the cost of meeting each person's needs with shared meals shows that transfers calculated based on individualized diets are unlikely to be sufficient to meet the needs of all household members. Cash transfers with amounts exceeding those of individualized diets coupled with nutritional behavioral change communication to ensure nutrient-rich foods are better targeted to meet individual needs, could then go some way to make social protection programs more nutrition-sensitive in a cost-effective manner (Hoddinott, Ahmed and Roy 2018; Ruel and Alderman 2013; Olney et al. 2021). Complementing social protection programs such as cash transfers with nutritional behavioral change communication is increasingly advocated with incipient evidence showing positive effects on individual nutritional outcomes for combined programs (though often not for cash transfers on their own) (Alderman, forthcoming). The estimated gap between individualized and shared dietary costs as well as local knowledge on meal sharing practices can provide further guidance on the extra amount of cash needed beyond the individualized dietary costs.

While filling an important gap arising from intrahousehold food allocation practices, several questions for future research on the ability of food systems to deliver nutrient adequacy also remain. Regularly collected 'main town' market prices may not reflect more remote local conditions nor the option to supplement market availability with own production (including through gardening). Future research should investigate how unit costs from household surveys could be used to estimate more precise least-cost diets, their composition, and the relative contribution of own production. Other modeling tools such as stochastic optimization could also be used to capture variation in the food environment and incorporate uncertainty in the underlying data.

**Acknowledgements**

We sincerely thank the two three anonymous referees and editor Timothy Richards for their invaluable feedback, insights, and suggestions while the manuscript was under review. The paper benefitted greatly from their contributions. We are also grateful to collaborators Yan Bai, Anna Herforth, and Stevier Kaiyatsa for their thougthful input and feedback throughout the process that culminated in this paper. This paper was part of the Changing Access to Nutritious Diets in Africa and South Asia (CANDASA) and Food Prices for Nutrition projects, supported by the Foreign Commonwealth Development Office (FCDO) of the United Kingdom and the Bill & Melinda Gates Foundation (INV-009668 and INV-016158).

**Data and code availability**

Replication code and supporting datasets are available on Harvard Dataverse at https://doi.org/10.7910/DVN/AXDJZ8. Code and data to replicate the datasets used in this analysis from raw sources (household survey, food composition, nutrient requirements, prices) are available on the author's GitHub at https://github.com/KateSchneider-FoodPol/Kate-Schneider-MalawiDietQualCostHHs-2021.

**Supplementary Materials**

Table A. Markets in CPI Price Monitoring Dataset Observed in IHPS Dataset

| Region | District | Market |
|---|---|---|
| North | Chitipa | Chitipa Boma |
| | Karonga | Karonga Boma |
| | Nkhatabay | Nkhatabay Boma |
| | Rumphi | Rumphi Boma |
| | Mzimba | Ekwendeni |
| Central | Mzimba | Ekwendeni |
| | Kasungu | Kasungu Boma |
| | Nkotakota | Nkhotakota Boma |
| | Ntchisi | Mponera |
| | Dowa | Mponera |
| | Salima | Salima Boma |
| | Lilongwe Non-City | Mitundu |
| | Mchinji | Mchinji Boma |
| | Dedza | Dedza Boma |
| | Ntcheu | Ntcheu Boma |
| South | Mangochi | Mangochi Boma |
| | Machinga | Liwonde |
| | Zomba Non-City | Jali |
| | Chiradzulu | Mbulumbuzi |
| | Blantyre Non-City | Lunzu |
| | Mwanza | Mwanza Boma |
| | Thyolo | Thyolo |
| | Mulanje | Chitakale |
| | Phalombe | Phalombe Boma |
| | Chikwawa | Nchalo |
| | Nsanje | Nsanje Boma |
| | Balaka | Balaka Boma |

Table B-1. Difference between market unit prices and reported unit costs, by size of difference

| Food item | Mean unit cost | Mean unit market price | Diff. (Market unit price - unit cost) | SE (Diff.) | p (Diff.) | | Degrees of freedom |
|---|---|---|---|---|---|---|---|
| Sun Dried fish (Large Variety) | 2,137 | 5,320 | 3,183.61 | 413.73 | 0.00000 | *** | 43 |
| Sun Dried fish (Medium Variety) | 1,793 | 3,875 | 2,082.35 | 106.28 | 0.00000 | *** | 177 |
| Goat | 1,103 | 3,063 | 1,960.49 | 628.87 | 0.00189 | ** | 753 |
| Beef | 1,178 | 2,787 | 1,608.85 | 789.67 | 0.04232 | * | 374 |
| Sun Dried fish (Small Variety) | 2,015 | 3,150 | 1,134.60 | 44.62 | 0.00000 | *** | 803 |
| Pork | 1,098 | 2,146 | 1,048.04 | 613.15 | 0.08837 | | 319 |
| Cooking oil | 970 | 1,391 | 421.14 | 16.78 | 0.00000 | *** | 2,282 |
| Maize ufa mgaiwa (normal flour) | 166 | 455 | 288.47 | 6.94 | 0.00000 | *** | 1,092 |
| Fresh milk | 365 | 631 | 266.06 | 137.12 | 0.05321 | | 326 |
| Maize ufa madeya (bran flour) | 156 | 393 | 237.69 | 30.91 | 0.00000 | *** | 43 |
| Buns, scones | 572 | 747 | 175.23 | 29.98 | 0.00000 | *** | 843 |
| Chicken | 1,326 | 1,490 | 164.43 | 35.63 | 0.00001 | *** | 168 |
| Onion | 328 | 487 | 159.51 | 7.79 | 0.00000 | *** | 2,125 |
| Pigeonpeas | 400 | 553 | 153.42 | 18.35 | 0.00000 | *** | 257 |
| Brown beans | 491 | 610 | 118.84 | 6.10 | 0.00000 | *** | 1,591 |
| Guava | 108 | 223 | 114.45 | 31.49 | 0.00071 | *** | 45 |
| Papaya | 76 | 184 | 107.27 | 19.22 | 0.00001 | *** | 24 |
| Groundnut | 275 | 378 | 103.10 | 8.53 | 0.00000 | *** | 542 |
| Cucumber | 187 | 270 | 82.31 | 32.24 | 0.01458 | * | 40 |
| Avocado | 142 | 217 | 75.62 | 12.29 | 0.00000 | *** | 69 |
| White beans | 424 | 493 | 69.40 | 10.75 | 0.00000 | *** | 472 |
| Tomato | 217 | 284 | 66.91 | 2.89 | 0.00000 | *** | 3,897 |
| Citrus | 122 | 189 | 66.22 | 8.71 | 0.00000 | *** | 144 |
| Cowpeas | 360 | 423 | 63.00 | 13.40 | 0.00000 | *** | 244 |
| Cassava tubers | 90 | 141 | 50.84 | 2.94 | 0.00000 | *** | 688 |
| Rice | 416 | 458 | 42.92 | 3.95 | 0.00000 | *** | 1,052 |
| Pumpkin | 52 | 87 | 34.73 | 7.57 | 0.00005 | *** | 36 |
| White sweet potato | 92 | 113 | 21.35 | 2.02 | 0.00000 | *** | 1,296 |
| Mango | 128 | 142 | 14.24 | 4.93 | 0.00435 | ** | 190 |
| Banana | 152 | 164 | 11.97 | 2.83 | 0.00003 | *** | 1,164 |

| Food item | Mean unit cost | Mean unit market price | Diff. (Market unit price - unit cost) | SE (Diff.) | p (Diff.) | | Degrees of freedom |
|---|---|---|---|---|---|---|---|
| Irish potato | 218 | 213 | -4.74 | 5.60 | 0.39800 | | 445 |
| Cabbage | 91 | 80 | -10.37 | 1.50 | 0.00000 | *** | 849 |
| Maize grain | 193 | 180 | -13.48 | 22.95 | 0.55971 | | 49 |
| Chinese cabbage | 230 | 190 | -39.61 | 10.78 | 0.00028 | *** | 348 |
| Mandazi | 708 | 665 | -42.73 | 9.90 | 0.00002 | *** | 1,193 |
| Eggs | 1,071 | 1,026 | -45.88 | 14.88 | 0.00209 | ** | 1,205 |
| Biscuits | 675 | 617 | -57.54 | 63.06 | 0.36230 | | 277 |
| Rape | 185 | 127 | -57.72 | 2.43 | 0.00000 | *** | 2,844 |
| Pumpkin leaves | 263 | 204 | -59.82 | 5.03 | 0.00000 | *** | 1,228 |
| Okra | 364 | 256 | -108.35 | 9.31 | 0.00000 | *** | 726 |
| Salt | 373 | 232 | -140.63 | 6.80 | 0.00000 | *** | 4,109 |
| Sugar | 671 | 394 | -277.03 | 20.34 | 0.00000 | *** | 2,355 |
| Powdered milk | 7,187 | 1,675 | -5,512.18 | 3,101.76 | 0.07746 | | 159 |
| Bread | 151,944 | 355 | -151,588.65 | 5,917.13 | 0.00000 | *** | 843 |

Notes: Unit costs and unit prices compared in the same month and year in nominal MWK. Extreme unit costs above the 99th percentile excluded. Red font indicates unit costs reported by households exceed market unit prices. *** p<0.001 ** p<0.01 *p<0.05

Table B-2. Difference Between Market Unit Prices and Reported Unit Costs, by Food Group

| | Food item | Mean unit cost | Mean unit market price | Diff. (Market unit price - unit cost) | SE (Diff.) | p (Diff.) | | Degrees of freedom |
|---|---|---|---|---|---|---|---|---|
| Staples | Bread | 151,944 | 355 | -151,588.65 | 5,917.13 | 0.00000 | *** | 843 |
| | Cassava tubers | 90 | 141 | 50.84 | 2.94 | 0.00000 | *** | 688 |
| | Irish potato | 218 | 213 | -4.74 | 5.60 | 0.39800 | | 445 |
| | Maize grain | 193 | 180 | -13.48 | 22.95 | 0.55971 | | 49 |
| | Maize ufa madeya (bran flour) | 156 | 393 | 237.69 | 30.91 | 0.00000 | *** | 43 |
| | Maize ufa mgaiwa (normal flour) | 166 | 455 | 288.47 | 6.94 | 0.00000 | *** | 1,092 |
| | Rice | 416 | 458 | 42.92 | 3.95 | 0.00000 | *** | 1,052 |
| | White sweet potato | 92 | 113 | 21.35 | 2.02 | 0.00000 | *** | 1,296 |
| Legumes | Brown beans | 491 | 610 | 118.84 | 6.10 | 0.00000 | *** | 1,591 |
| | Cowpeas | 360 | 423 | 63.00 | 13.40 | 0.00000 | *** | 244 |
| | Groundnuts | 275 | 378 | 103.10 | 8.53 | 0.00000 | *** | 542 |
| | Pigeonpea | 400 | 553 | 153.42 | 18.35 | 0.00000 | *** | 257 |
| | White beans | 424 | 493 | 69.40 | 10.75 | 0.00000 | *** | 472 |
| Vegetables | Cabbage | 91 | 80 | -10.37 | 1.50 | 0.00000 | *** | 849 |
| | Chinese cabbage | 230 | 190 | -39.61 | 10.78 | 0.00028 | *** | 348 |
| | Cucumber | 187 | 270 | 82.31 | 32.24 | 0.01458 | * | 40 |
| | Okra | 364 | 256 | -108.35 | 9.31 | 0.00000 | *** | 726 |
| | Onion | 328 | 487 | 159.51 | 7.79 | 0.00000 | *** | 2,125 |
| | Pumpkin | 52 | 87 | 34.73 | 7.57 | 0.00005 | *** | 36 |
| | Pumpkin leaves | 263 | 204 | -59.82 | 5.03 | 0.00000 | *** | 1,228 |
| | Rape | 185 | 127 | -57.72 | 2.43 | 0.00000 | *** | 2,844 |
| | Tomato | 217 | 284 | 66.91 | 2.89 | 0.00000 | *** | 3,897 |
| Fruits | Avocado | 142 | 217 | 75.62 | 12.29 | 0.00000 | *** | 69 |
| | Banana | 152 | 164 | 11.97 | 2.83 | 0.00003 | *** | 1,164 |
| | Citrus | 122 | 189 | 66.22 | 8.71 | 0.00000 | *** | 144 |
| | Guava | 108 | 223 | 114.45 | 31.49 | 0.00071 | *** | 45 |
| | Mango | 128 | 142 | 14.24 | 4.93 | 0.00435 | ** | 190 |
| | Papaya | 76 | 184 | 107.27 | 19.22 | 0.00001 | *** | 24 |

|  | Food item | Mean unit cost | Mean unit market price | Diff. (Market unit price – unit cost) | SE (Diff.) | p (Diff.) |  | Degrees of freedom |
|---|---|---|---|---|---|---|---|---|
| **Animal-source foods** | Beef | 1,178 | 2,787 | 1,608.85 | 789.67 | 0.04232 | * | 374 |
|  | Chicken | 1,326 | 1,490 | 164.43 | 35.63 | 0.00001 | *** | 168 |
|  | Eggs | 1,071 | 1,026 | -45.88 | 14.88 | 0.00209 | ** | 1,205 |
|  | Fresh milk | 365 | 631 | 266.06 | 137.12 | 0.05321 |  | 326 |
|  | Goat | 1,103 | 3,063 | 1,960.49 | 628.87 | 0.00189 | ** | 753 |
|  | Pork | 1,098 | 2,146 | 1,048.04 | 613.15 | 0.08837 |  | 319 |
|  | Powdered milk | 7,187 | 1,675 | -5,512.18 | 3,101.76 | 0.07746 |  | 159 |
|  | Sun Dried fish (Large Variety) | 2,137 | 5,320 | 3,183.61 | 413.73 | 0.00000 | *** | 43 |
|  | Sun Dried fish (Medium Variety) | 1,793 | 3,875 | 2,082.35 | 106.28 | 0.00000 | *** | 177 |
|  | Sun Dried fish (Small Variety) | 2,015 | 3,150 | 1,134.60 | 44.62 | 0.00000 | *** | 803 |
|  | Cooking oil | 970 | 1,391 | 421.14 | 16.78 | 0.00000 | *** | 2,282 |
|  | Salt | 373 | 232 | -140.63 | 6.80 | 0.00000 | *** | 4,109 |
| **Sweets** | Biscuits | 675 | 617 | -57.54 | 63.06 | 0.36230 |  | 277 |
|  | Buns, scones | 572 | 747 | 175.23 | 29.98 | 0.00000 | *** | 843 |
|  | Mandazi | 708 | 665 | -42.73 | 9.90 | 0.00002 | *** | 1,193 |
|  | Sugar | 671 | 394 | -277.03 | 20.34 | 0.00000 | *** | 2,355 |

Table C. Food Items by Food Group in Price Dataset

| Food Group | Items | Food Group | Items |
|---|---|---|---|
| Cereals & Cereal Products | Maize flour (dehulled) | Vitamin-A rich fruits | Mangoes |
| | Maize flour (whole grain) | | Oranges |
| | Maize grain | | Papaya |
| | Maize grain, Admarc | | Tomatoes |
| | Rice grain | Vit-A rich Vegetables | Pumpkin |
| | White bread | Other Fruits | Avocado |
| Dark Green Leafy Vegetables | Chinese cabbage | | Banana |
| | Pumpkin leaves | | Guava |
| | Rape leaves | Other Vegetables | Okra |
| Eggs | Chicken eggs | | Onions |
| Fish & Seafood | Cichlid (Utaka, dried) | | Cabbage |
| | Oreochromis lidole, dry[a] | | Cucumber |
| | Oreochromis lidole, fresh[a] | | Eggplant |
| | Sardine (Usipa, sun dried) | | Green beans |
| Flesh Meat | Beef | Roots & Tubers | Cassava |
| | Goat | | Irish potatoes |
| | Live chicken | | Sweet potatoes |
| | Pork | Salty & fried foods | Mandazi |
| Legumes | Brown beans | Sweets & Confectionary | Biscuits |
| | Cowpeas | | Brown sugar |
| | Groundnuts | | White buns |
| | Pigeon peas | | White sugar |
| | White beans | Stimulants, Spices, & Condiments[b] | Salt |
| Milk & Milk Products | Fresh milk | | |
| | Powdered milk | Caloric beverages[b] | Coca-cola |
| Oils & Fats | Cooking oil | | |
| | Cooking oil refill | Total items (N) | 51 |

[a] Tilapia, known locally as *chambo*.
[b] The food list also monitors the price of three types of tea and a fermented maize-based drink, *Maheu*. Tea is excluded because it confers no essential nutrients. *Maheu* has been excluded from the analysis for lack of food composition data.

Table D. Nutrient Composition and Density by Food Item and Nutrient

| Nutrient | Items with highest nutrient quantity per 100g edible portion[a] | Items with highest nutrient density (quantity per unit energy)[a] |
|---|---|---|
| Energy | Cooking oil, Groundnuts, Powdered milk, Biscuits, Sugar, Maize flour, Pigeon peas, Dry *Usipa*, Cowpeas, Rice | Cooking oil, Groundnuts, Powdered milk, Biscuits, Sugar, Maize flour, Pigeon peas, Dried *Usipa*, Cowpeas, Rice |
| Carbohydrate | Sugar, Rice, Maize flour. Maize grain, Biscuits, Pigeon peas, Cowpeas, White beans, Brown beans, White bread | Coca-cola, Sugar, Cucumber, Cassava, Mango, Banana, Sweet potato, Oranges, Rice, Papaya |
| Protein | Dry *Chambo*, Dry *Usipa*, *Utaka*, Powdered milk, Brown beans, Groundnuts, Cowpeas, Pigeon peas, White beans, Chicken | Dry *Chambo*, Beef, Dry *Usipa*, Chicken, Fresh *Chambo*, *Utaka*, Goat, Eggs, Pumpkin leaves, Brown beans, Pork |
| Lipids | Cooking oil, Groundnuts, Powdered milk, Pork, Biscuits, *Utaka*, Avocado, Goat, Eggs, Dry *Usipa* | Cooking oil, Avocado, Pork, Groundnuts, Eggs, Goat, Fresh milk, Powdered milk, *Utaka*, Biscuits |
| Vitamin A[b] | Rape leaves, Powdered milk, Pumpkin, Biscuits, Pumpkin leaves, Mangoes | Rape leaves, Pumpkin leaves, Pumpkin, Chinese cabbage, Mangoes, Tomatoes |
| Retinol | Powdered milk, Chicken, Biscuits, Eggs, Fresh milk | Chicken, Eggs, Fresh milk, Powdered milk, Biscuits |
| Vitamin C | Guava, Papaya, Rape leaves, Oranges, Okra, Chinese cabbage, Cassava, Cabbage, Mangoes, Pumpkin leaves | Guava, Chinese cabbage, Papaya, Rape leaves, Oranges, Cabbage, Pumpkin leaves, Okra, Tomatoes |
| Vitamin E | Cooking oil, Groundnuts | Pumpkin leaves, Rape leaves, Cooking oil, Pumpkin, Tomatoes, Groundnuts, Papaya, Mangoes, Guava |
| Thiamin | Groundnuts, White beans, Pork, Cowpeas, Pigeon peas, Brown beans, White buns, Maize grain, Maize flour | Pork, Irish potatoes, White beans, Cowpeas, White buns, Rape leaves, Green beans, Cucumber, Pumpkin leaves |
| Riboflavin | Powdered milk, Dry *Usipa*, Eggs, Goat, Dry *Chambo*, Brown beans, Pork, White beans, Beef, Pigeon peas | Powdered milk, Eggs, Rape leaves, Pumpkin leaves, Fresh milk, Cucumber, Beef, Okra, Dried *Usipa*, Goat |
| Niacin | Dry Usipa, Groundnuts, Beef, Goat, Pork, Chicken, Dry *Chambo*, Cowpeas, Pigeon peas, Maize grain | Dried *Usipa*, Beef, Goat, Chicken, Groundnuts, Chinese cabbage, Tomatoes, Green beans, Irish potatoes, Pumpkin leaves |

| Nutrient | Items with highest nutrient quantity per 100g edible portion[a] | Items with highest nutrient density (quantity per unit energy)[a] |
|---|---|---|
| Vitamin B6 | Dry *Usipa* | Guava, Dried *Usipa*, Rape leaves, Okra, Pumpkin leaves, Banana, Irish potatoes, Tomatoes, Onions, Cucumber |
| Folate | Cowpeas, Brown beans, White beans, Pigeon peas, Rape leaves, Okra, Groundnuts | Rape leaves, Okra, Cowpeas, Pumpkin leaves, Brown beans, White beans, Pigeon peas |
| Vitamin B12 | Dry *Usipa*, Dry *Chambo*, Eggs, Powdered milk, Beef | Dried *Usipa*, Dry *Chambo*, Beef, Eggs, Goat, Fresh milk, Powdered milk, Pork, Chicken |
| Calcium | Rape leaves, Pumpkin leaves, *Utaka*, Dry *Chambo*, Dry *Usipa*, Cabbage, Papaya, Powdered milk | Pumpkin leaves, Rape leaves, Cabbage, Papaya, Tomatoes, Onions, *Utaka*, Dry *Chambo*, Chinese cabbage, Dried *Usipa* |
| Copper | Tomatoes, Cabbage, Papaya, Sweet potatoes, Pigeon peas, Onions, Pumpkin leaves, Groundnuts, Rape leaves, Cowpeas | Tomatoes, Cabbage, Pumpkin leaves, Papaya, Onions, Rape leaves, Sweet potatoes, Mangoes |
| Iron | Pumpkin leaves, Dry *Chambo*, Cabbage, Rape leaves, *Utaka* | Pumpkin leaves, Cabbage, Rape leaves, Tomatoes, Dry *Chambo*, Onions, Papaya, Beef, *Utaka* |
| Magnesium | Pumpkin leaves, *Utaka*, Papaya, Rape leaves, Cabbage | Pumpkin leaves, Cabbage, Rape leaves, Papaya, Onions |
| Phosphorus | Dry *Usipa*, Dry *Chambo*, Powdered milk, White beans, Brown beans, Cowpeas, Groundnuts, Pigeon peas, Rice, Maize grain | Dried *Usipa*, Dry *Chambo*, Pumpkin leaves, Beef, Cucumber, Okra, Powdered milk, Fresh milk, White beans, Eggs |
| Selenium | White beans, Pumpkin leaves, Papaya, Brown beans, Tomatoes, Rape leaves, Cabbage, Cowpeas | Pumpkin leaves, Tomatoes, Cabbage, Papaya, Rape leaves, Onions, Mangoes, White beans, Brown beans |
| Zinc | Dry *Usipa*, Rape leaves, Pumpkin leaves, Dry *Chambo*, Onions, Pork, Cabbage, Goat, Powdered milk, Brown beans | Rape leaves, Pumpkin leaves, Tomatoes, Cabbage, Onions, Dried *Usipa*, Chinese cabbage, Papaya, Goat, Beef |
| Sodium | Salt, White bread, White buns, Biscuits, Powdered milk, Dry *Chambo*, Dry *Usipa* | White bread, White buns |

[a] Listed in descending order of quantity or density. Listing top sources where a natural divide in density or quantity occurs, otherwise top 10 items listed.
[b] Sugar and cooking oil are fortified with vitamin A in Malawi.

Table E. Percent of Nutrients and Expenditure Supplied by Food Items Included in the Retail Market Food Price List

| % Consumption from items in food price list | 2013 | | 2016/17 | | Overall | |
|---|---|---|---|---|---|---|
| | Mean | (SE) | Mean | (SE) | Mean | (SE) |
| Energy | 94.33 | (0.553) | 94.11 | (0.454) | 94.19 | (0.389) |
| Carbohydrate | 94.36 | (0.580) | 94.65 | (0.453) | 94.54 | (0.401) |
| Protein | 93.28 | (0.635) | 92.43 | (0.622) | 92.75 | (0.507) |
| Lipids | 95.20 | (0.542) | 94.36 | (0.443) | 94.67 | (0.382) |
| **Vitamin A** | **78.52** | **(1.705)** | **84.27** | **(1.571)** | **82.13** | **(1.303)** |
| **Vitamin C** | **84.63** | **(1.072)** | **86.26** | **(1.038)** | **85.65** | **(0.935)** |
| Vitamin E | 96.04 | (0.341) | 96.10 | (0.313) | 96.08 | (0.265) |
| Thiamin | 94.91 | (0.574) | 94.63 | (0.513) | 94.73 | (0.433) |
| Riboflavin | 91.47 | (0.654) | 92.91 | (0.504) | 92.38 | (0.462) |
| Niacin | 92.28 | (0.670) | 92.56 | (0.541) | 92.45 | (0.493) |
| Vitamin B6 | 90.24 | (0.772) | 91.29 | (0.585) | 90.90 | (0.515) |
| Folate | 92.03 | (0.646) | 90.68 | (0.798) | 91.18 | (0.625) |
| **Vitamin B12** | **83.17** | **(1.983)** | **94.45** | **(0.832)** | **90.06** | **(1.068)** |
| **Calcium** | **84.03** | **(1.161)** | **89.89** | **(0.607)** | **87.71** | **(0.673)** |
| Copper | 95.35 | (0.485) | 94.55 | (0.489) | 94.85 | (0.425) |
| **Iron** | **86.31** | **(0.992)** | **92.16** | **(0.527)** | **89.98** | **(0.555)** |
| Magnesium | 90.67 | (0.723) | 92.53 | (0.527) | 91.84 | (0.488) |
| **Phosphorus** | **89.56** | **(0.905)** | **89.28** | **(0.838)** | **89.38** | **(0.733)** |
| Selenium | 90.49 | (1.012) | 89.90 | (1.057) | 90.12 | (0.926) |
| Zinc | 92.69 | (0.578) | 92.80 | (0.518) | 92.76 | (0.442) |
| Sodium | 96.79 | (0.394) | 97.74 | (0.262) | 97.38 | (0.223) |
| **Total Expenditure** | 90.01 | (0.650) | 89.86 | (0.625) | 89.91 | (0.512) |

Notes: Population statistics corrected using sampling weights. Heteroskedasticity robust standard errors clustered at the enumeration area level.

Table F. Model Fit Statistics

| | Stochastic Trend Dummy Model | Trigonometric Model |
|---|---|---|
| *Individualized Diets* | | |
| (N=66,794) | | |
| F-statistic | $F_{11,98}$=14.97*** | $F_{2,98}$=5.252*** |
| Adj. R-squared | 0.0116 | 0.0009 |
| AIC | 0.8536 | 0.8642 |
| BIC | 0.8552 | 0.8646 |
| *Household Sharing* | | |
| (N=40,067) | | |
| F-statistic | $F_{11,98}$=17.20*** | $F_{2,98}$=15.36*** |
| Adj. R-squared | 0.0189 | 0.0026 |
| AIC | 0.9077 | 0.9240 |
| BIC | 0.9103 | 0.9246 |

Notes: Population statistics corrected using sampling weights. Preferred specification in bold. AIC and BIC are reported on a per observation basis. Heteroskedasticity robust standard errors clustered at the enumeration area level in all specifications. *$p < 0.05$ **$p < 0.01$ ***$p < 0.001$

Table G. Individual Daily Cost of Nutrient Adequacy over 25 markets January 2013-July 2017, All Individual Types by Nutrient Requirement Group

| | Population Share % | Months with Solution (%) Mean | (SD) | Cost/day (2011 US$) Median | (SD) |
|---|---|---|---|---|---|
| Infant (all) 6 months-1 y | 1.35 | **80.00** | (40.01) | 0.08 | (0.03) |
| Child (all) 1-2 y[a] | 5.45 | **62.36** | (48.46) | 3.18 | (11.15) |
| Child (M) 3 y | 1.57 | **86.61** | (34.06) | 1.43 | (3.74) |
| Child (F) 3 y | 1.82 | **86.46** | (34.22) | 1.35 | (4.62) |
| Child (M) 4-8 y | 8.15 | 99.06 | (9.68) | 1.14 | (0.40) |
| Child (F) 4-8 y | 8.46 | 99.02 | (9.83) | 0.95 | (0.38) |
| Adolescent (M) 9-13 y | 7.92 | 97.76 | (14.79) | 1.78 | (0.60) |
| Adolescent (M) 14-18 y | 5.91 | 97.13 | (16.69) | 2.57 | (2.44) |
| Adult (M) 19-30 y | 8.14 | 97.15 | (16.64) | 2.57 | (2.10) |
| Adult (M) 31-50 y | 8.19 | 96.96 | (17.17) | 2.57 | (2.09) |
| Adult (M) 51-70 y | 3.04 | **91.37** | (28.08) | 2.47 | (10.22) |
| Older Adult (M) 70+ y | 0.99 | **82.68** | (37.85) | 2.29 | (13.85) |
| Adolescent (F) 9-13 y | 7.76 | 97.32 | (16.14) | 1.44 | (0.68) |
| Adolescent (F) 14-18 y | 5.53 | 96.85 | (17.47) | 1.94 | (0.70) |
| Adult (F) 19-30 y | 6.84 | 97.23 | (16.42) | 2.04 | (0.96) |
| Adult (F) 31-50 y | 7.31 | 96.98 | (17.13) | 2.00 | (0.95) |
| Adult (F) 51-70 y | 3.58 | 93.23 | (25.13) | 2.07 | (4.33) |
| Older Adult (F) 70+ y | 1.25 | **87.65** | (32.90) | 2.01 | (7.07) |
| Lactation (F) 14-18 y | 0.28 | **56.79** | (49.54) | 2.76 | (1.63) |
| Lactation (F) 19-30 y | 3.41 | **57.04** | (49.51) | 2.76 | (1.87) |
| Lactation (F) 31-50 y | 1.64 | **56.94** | (49.52) | 2.76 | (2.03) |
| **Population weighted Average** | | 93.06 | | 2.38 | |

Notes: Population shares calculated with survey weights from household data. Age-sex groups based on *Dietary Reference Intakes* categories, disaggregating 3-year-old children from the micronutrient group aged 1-3 years to accommodate separate estimated energy requirement equations.

[a] Upper bound of protein AMDR is relaxed (increased) by 50% for children 6-35 months.